# TBA Equations of 1D Hubbard Model and High-Temperature Expansion

Minoru Takahashi and Masahiro Shiroishi

*Institute for Solid State Physics, University of Tokyo,*

*Kashiwanoha 5-1-5, Kashiwa, Chiba, 277-8581 Japan*

(Dated: October 31, 2001)

## Abstract

New numerical method to calculate thermodynmic Bethe ansatz equations is proposed based on Newton's method. Thermodynamic quantities of one-dimensional Hubbard model is numerically calculated and compared with high temperature expansion and numerical results of quantum transfer matrix method by Jüttner, Klümper and Suzuki. The coincidence is surprisingly good. We get high-temperature expansion of grand potential up to $\beta^6$.

## I. INTRODUCTION

Many years ago, one of the authors (MT) proposed thermodynamic Bethe ansatz (TBA) equations for one-dimensional Hubbard model [2]. In this theory several kind of strings are assumed and it is widely believed that this set of equations give the exact thermodynamic quantities of this model. Low-temperature thermodynamics were investigated by MT [3] and actual numerical calculations at finite temperature were done by Kawakami, Usuki and Okiji [4]. Essler, Korepin and Schoutens [5] counted the number of states by the single $k$ excitations, $\Lambda$ strings and $k - \Lambda$ strings. They found that total number of these Bethe ansatz states and their relatives is $4^{N_a}$, where $N_a$ is the length of the systems. This implies that the Bethe ansatz can give all eienstates and eigenvalues. However some physicists are still skeptical for this theory [8]. Recently Charret et al [7] did the numerical calculation of this equation and concluded that it does not coincide with high temperature expansion and quantum transfer matrix(QTM) method by Jüttner, Klümper and Suzuki [6]. In this paper we give a practical method to calculate numerically TBA equations which has infinite unknown functions. Numerical results completely coincide with those of QTM and HTE. Charret et al's numerical calculation of TBA equations is wrong. The Hubbard Hamiltonian is

$$\mathcal{H}(t, U, A, h) =$$
$$-t \sum_{<ij>} \sum_{\sigma} (c_{i\sigma}^\dagger c_{j\sigma} + c_{j\sigma}^\dagger c_{i\sigma}) + U \sum_{i=1}^{N_a} c_{i\uparrow}^\dagger c_{i\uparrow} c_{i\downarrow}^\dagger c_{i\downarrow}$$
$$-A \sum_{i=1}^{N_a} (c_{i\uparrow}^\dagger c_{i\uparrow} + c_{i\downarrow}^\dagger c_{i\downarrow}) - h \sum_{i=1}^{N_a} (c_{i\uparrow}^\dagger c_{i\uparrow} - c_{i\downarrow}^\dagger c_{i\downarrow}). \qquad (1)$$

Here $c_{j\sigma}^\dagger$ and $c_{j\sigma}$ are creation and annihilation operators of an electron at site $j$. $<ij>$ means that sites $i$ and $j$ are nearest neighbors. $N_a$ is the number of atoms. We put $t > 0, U > 0$. Thermodynamic potential per site $g$ at temperature $T$ is determined by

$$g = e_0 - A - T \Big\{ \int_{-\pi}^{\pi} \rho_0(k) \ln(1 + \zeta(k)) dk$$
$$+ \int_{-\infty}^{\infty} \sigma_0(\Lambda) \ln(1 + \eta_1(\Lambda)) d\Lambda \Big\}. \qquad (2)$$

Here $e_0, \rho_0(k)$ and $\sigma_0(\Lambda)$ are energy per cite, distribution functions of $k$'s and $\Lambda$'s at $T = h = U/2 - A = 0$, (half-filled, zero-field ground state),

$$e_0 = -4t \int_0^\infty \frac{J_0(\omega)J_1(\omega)d\omega}{\omega(1+\exp(2U'\omega))}, \tag{3}$$

$$\sigma_0(\Lambda) = \int_{-\pi}^{\pi} s(\Lambda - \sin k)\frac{dk}{2\pi}, \tag{4}$$

$$\rho_0(k) = \frac{1}{2\pi} + \cos k \int_{-\infty}^{\infty} a_1(\Lambda - \sin k)\sigma_0(\Lambda)d\Lambda, \tag{5}$$

and

$$a_1(x) \equiv \frac{U'}{\pi(U'^2 + x^2)}, \quad s(x) \equiv \frac{1}{4U'}\mathrm{sech}\frac{\pi x}{2U'}, \quad U' \equiv \frac{U}{4t}. \tag{6}$$

$\zeta(k)$ and $\eta_1(\Lambda)$ are hole-particle ratios of $k$ excitations and single $\Lambda$ excitations. These are determined by thermodynamic Bethe ansatz equations for $k$ excitations, $\Lambda$ strings and $k-\Lambda$ strings;

$$\ln \zeta(k) = \frac{\kappa_0(k)}{T}$$
$$+ \int_{-\infty}^{\infty} d\Lambda s(\Lambda - \sin k) \ln\left(\frac{1+\eta_1'(\Lambda)}{1+\eta_1(\Lambda)}\right), \tag{7}$$

$$\ln \eta_1(\Lambda) = s * \ln(1+\eta_2(\Lambda))$$
$$- \int_{-\pi}^{\pi} s(\Lambda - \sin k)\ln(1+\zeta^{-1}(k))\cos k dk, \tag{8}$$

$$\ln \eta_1'(\Lambda) = s * \ln(1+\eta_2'(\Lambda))$$
$$- \int_{-\pi}^{\pi} s(\Lambda - \sin k)\ln(1+\zeta(k))\cos k dk, \tag{9}$$

$$\ln \eta_j(\Lambda) = s * \ln\{(1+\eta_{j-1}(\Lambda))(1+\eta_{j+1}(\Lambda))\},$$
$$j \geq 2, \tag{10}$$

$$\ln \eta_j'(\Lambda) = s * \ln\{(1+\eta_{j-1}'(\Lambda))(1+\eta_{j+1}'(\Lambda))\},$$
$$j \geq 2, \tag{11}$$

$$\lim_{n\to\infty} \frac{\ln \eta_n(\Lambda)}{n} = \frac{2h}{T}, \tag{12}$$

$$\lim_{n\to\infty} \frac{\ln \eta_n'(\Lambda)}{n} = \frac{U-2A}{T}. \tag{13}$$

Here $s * f(\Lambda) \equiv \int_{-\infty}^{\infty} s(\Lambda - \Lambda')f(\Lambda')d\Lambda'$ and $\kappa_0(k)$ is defined by

$$\kappa_0(k) \equiv -2t\cos k$$
$$-4t \int_{-\infty}^{\infty} d\Lambda s(\Lambda - \sin k)\Re\sqrt{1-(\Lambda-U'i)^2}. \tag{14}$$

Details of derivation are given in references [1, 2, 9, 10]. Noting that $\ln(1 + \zeta^{-1}) = \ln(1 + \zeta) - \ln \zeta$ in (8) and substituting (7) we have

$$\ln \eta_1(\Lambda) = \frac{s_1(\Lambda)}{T} + s * \ln(1 + \eta_2(\Lambda))$$
$$- \int_{-\pi}^{\pi} s(\Lambda - \sin k) \ln(1 + \zeta(k)) \cos k dk,$$
$$s_1(\Lambda) \equiv -2t \int_{-\pi}^{\pi} \cos^2 k s(\Lambda - \sin k) dk. \qquad (15)$$

In Ref. [2] this set of equations were solved analytically in the limits $T \to 0$, $t \to 0$, and $U \to 0$ and coincided with known exact results. In a recent paper Charret et al [7] solved numerically this set of equations at high temperature and argued that there is discrepancy from high temperature expansion and numerical results of Jüttner, Klümper and Suzuki equations [6]. We recalculate the same quantities in this region and find that the results coincide with high temperature expansion and JKS equations in high accuracy. In III we review $t$ expansion for one-dimensional Hubbard model by the conventional linked cluster expansion. From the $t$ expansion, we can derive $\beta$ expansion of $-g\beta$ for the 1D Hubbard model up to $\beta^6$. Expansions of susceptibility and specific heat are obtained. In Appendix A, we can perform $t$ expansion of TBA equations. The results coincide with the cluster expansion up to the second order. We expect that the higher terms also coincide.

## II. TRUNCATION OF TBA EQUATIONS TO FINITE UNKNOWN FUNCTIONS

As an approximation we replace $s(\Lambda)$ by $\frac{1}{2}\delta(\Lambda)$ at $j > n_c$ in equations (10) and (11). Then we get the difference equations

$$\eta_j(\Lambda)^2 = (1 + \eta_{j-1}(\Lambda))(1 + \eta_{j+1}(\Lambda)),$$
$$\eta'_j(\Lambda)^2 = (1 + \eta'_{j-1}(\Lambda))(1 + \eta'_{j+1}(\Lambda)), \ j > n_c. \qquad (16)$$

This approximation is reasonable because functions $\eta_j(\Lambda)$ and $\eta'_j(\Lambda)$ vary very slowly at sufficiently large $j$ and $\int_{-\infty}^{\infty} s(\Lambda)d\Lambda = 1/2$. General solutions of these difference equations are

$$\eta_j(\Lambda) = \left(\frac{\sinh(f(\Lambda) + j)a}{\sinh a}\right)^2 - 1,$$
$$\eta'_j(\Lambda) = \left(\frac{\sinh(g(\Lambda) + j)b}{\sinh b}\right)^2 - 1, \ j \geq n_c. \qquad (17)$$

From the conditions (12) and (13), parameters $a$ and $b$ must be $h/T$ and $(U/2 - A)/T$. Then $1 + \eta_{n_c+1}(\Lambda)$ and $1 + \eta'_{n_c+1}(\Lambda)$ are represented by

$$\left(\cosh\frac{h}{T}\sqrt{1 + \eta_{n_c}(\Lambda)} + \sqrt{1 + \sinh^2\frac{h}{T}(1 + \eta_{n_c}(\Lambda))}\right)^2,$$

$$\left(\cosh\frac{h'}{T}\sqrt{1 + \eta'_{n_c}(\Lambda)} + \sqrt{1 + \sinh^2\frac{h'}{T}(1 + \eta'_{n_c}(\Lambda))}\right)^2,$$

$$h' \equiv U/2 - A. \tag{18}$$

Thus integral equations with infinite unknown functions are approximated by those with $2n_c + 1$ unknowns $\ln \eta'_{n_c}(\Lambda), ..., \ln \eta'_1(\Lambda), \ln \zeta(k), \ln \eta_1(\Lambda), ..., \ln \eta_{n_c}(\Lambda)$. Then equations to be solved are

$$z_1 = s * \ln[(1 + \exp z_2)(\cosh u_2\sqrt{1 + \exp z_1} + \sqrt{1 + \sinh^2 u_2(1 + \exp z_1)})^2],$$

$$z_j = s * \ln(1 + \exp z_{j-1})(1 + \exp z_{j+1}), \quad j = 2, ..., n_c - 1,$$

$$z_{n_c} = s * \ln(1 + \exp z_{n_c-1}) - \int_{-\pi}^{\pi} s(\Lambda - \sin k) \ln(1 + \exp z_{n_c+1}) \cos k \, dk,$$

$$z_{n_c+1} = u_1\kappa_0 + \int_{-\infty}^{\infty} d\Lambda s(\Lambda - \sin k) \ln\left(\frac{1 + \exp z_{n_c}}{1 + \exp z_{n_c+2}}\right),$$

$$z_{n_c+2} = u_1 s_1 + s * \ln(1 + \exp z_{n_c+3}) - \int_{-\pi}^{\pi} s(\Lambda - \sin k) \ln(1 + \exp z_{n_c+1}) \cos k \, dk,$$

$$z_j = s * \ln(1 + \exp z_{j-1})(1 + \exp z_{j+1}), \quad j = n_c + 3, ..., 2n_c,$$

$$z_{2n_c+1} = s * \ln[(1 + \exp z_{2n_c})(\cosh u_3\sqrt{1 + \exp z_{2n_c+1}} + \sqrt{1 + \sinh^2 u_3(1 + \exp z_{2n_c+1})})^2].$$

$$\tag{19}$$

We introduce three thermodynamic parameters:

$$u_1 \equiv 1/T, \quad u_2 \equiv (U/2 - A)/T, \quad u_3 \equiv h/T.$$

For actual numerical calculations we choose $L$ discrete points of $k$ and $\Lambda$ as follows:

$$k_j = \pi(j - 1/2)/L, \quad \Lambda_j = \sin q_j \sqrt{1 + \frac{U'^2}{\cos^2 q_j}},$$

$$q_j = \pi(j - 1/2)/(2L), \quad j = 1, ..., L. \tag{20}$$

Here function $\Lambda = \sin q\sqrt{1 + (U'/\cos q)^2}$ is the inverse function of $\int^{\Lambda} \Re(1 - (t - U'i)^2)^{-1/2} dt$. We think that this change of parameters is reasonable because the change of functions is very slow at large $\Lambda$. For very big $U'$ this function behaves as $U' \tan q$ and for small $U'$ it behaves as $\sin q$. Unknown functions are represented by vectors with length $L$ and integration kernels are represented by $L \times L$ matrices.

Usually this kind of non-linear equations is calculated by successive iterations, which is called Kepler's method. In the solution of TBA equations we need to repeat several tens or several hundreds times of iterations to get a good convergence.

So here we propose to use Newton's method. Consider a coupled non-linear equations:

$$X_j - F_j(X_1, X_2, ..., X_N) = 0, \quad j = 1, ..., N. \tag{21}$$

For approximate vectors $X_j^{(l)}$ assume that we have deviations $\Delta_j$:

$$X_j^{(l)} - F_j(X_1^{(l)}, X_2^{(l)}, ..., X_N^{(l)}) = \Delta_j. \tag{22}$$

In Kepler's method next approximation is

$$X_j^{(l+1)} = X_j^{(l)} + \Delta_j. \tag{23}$$

In Newton's method we put

$$X_j^{(l+1)} = X_j^{(l)} + \xi_j, \tag{24}$$

where $\xi_j$ is the solution of linear equation

$$\sum_j (\delta_{i,j} - \frac{\partial F_i(X_1^{(l)}, X_2^{(l)}, ..., X_N^{(l)})}{\partial X_j})\xi_j = \Delta_i. \tag{25}$$

This method is much faster than Kepler's method. But we must solve linear equations with $N \times N$ matrix. In our TBA problem, $N$ is $(2n_c + 1)L$. This large matrix is block tridiagonal. Regarding $L \times L$ blocks as a number we can solve this set of linear equations. We need only 5-6 times of iterations at most to get sufficient convergence $\sum_j |\Delta_j| < 10^{-8}$. We can get the thermodynamic potential through equation (2):

$$\begin{aligned}\frac{g}{T} &= (e_0 - \frac{U}{2})u_1 + u_2 \\ &- \int \rho_0(k) \ln(1 + \exp z_{n_c+1}(k)) dk \\ &- \int \sigma_0(\Lambda) \ln(1 + \exp z_{n_c+2}(\Lambda)) d\Lambda. \end{aligned} \tag{26}$$

To get the first order thermodynamic quantities like magnetization ($m$), electron density ($n$) and entropy we need to calculate $\partial(g/T)/\partial u_1$, $\partial(g/T)/\partial u_2$, $\partial(g/T)/\partial u_3$.

$$\partial_i \frac{g}{T} = (e_0 - \frac{U}{2})\delta_{1i} + \delta_{2i}$$
$$- \int \rho_0(k) \frac{\partial_i z_{n_c+1}(k)}{1 + \exp(-z_{n_c+1}(k))} dk$$
$$- \int \sigma_0(\Lambda) \frac{\partial_i z_{n_c+2}(\Lambda)}{1 + \exp(-z_{n_c+2}(\Lambda))} d\Lambda. \tag{27}$$

The equations for $\partial_i z_\alpha$ is a linear equation which has the same homogeneous term with that in Newton's method. Inhomogeneous terms are calculated from $z_\alpha$. Therefore we can calculate these quantities by one operation of linear calculation,

$$e = \partial_1(g/T) + A\partial_2(g/T),$$
$$n = \partial_2(g/T), \quad m = \partial_3(g/T),$$
$$\text{entropy} = u_1(e - g) - u_3 m + (u_2 - Uu_1/2)n. \tag{28}$$

To calculate the second order thermodynamic quantities such as specific heat, susceptibility and compressibility we need $3 \times 3$ tensor $\partial_i \partial_j(g/T)$. As this is a symmetric tensor, we need to calculate six components. We consider equations for

$$f_\alpha^{(ij)} = \partial_i \partial_j z_\alpha + \frac{1}{1 + \exp z_\alpha}(\partial_i z_\alpha)(\partial_j z_\alpha), \quad i \leq j. \tag{29}$$

Tensor is given by

$$\partial_i \partial_j \frac{g}{T} = - \int \rho_0(k) \frac{f_{n_c+1}^{(ij)}(k)}{1 + \exp(-z_{n_c+1}(k))} dk$$
$$- \int \sigma_0(\Lambda) \frac{f_{n_c+2}^{(ij)}(\Lambda)}{1 + \exp(-z_{n_c+2}(\Lambda))} d\Lambda. \tag{30}$$

The inhomogeneous terms are calculated from $z_\alpha$ and $\partial_i z_\alpha$. So we can calculate thermodynamic quantities from 10 quantities $g/T$ and its derivatives for given temperature, magnetic field and chemical potential after several times of linear equation solving. Specific heat $c$ in constant $n$ and $m$, susceptibility $\chi$ in constant $n$ and temperature, compressibility $\kappa$ in constant temperature and $m/n$ are given by

$$c = u_1^2(D_{11} + \frac{D_{13}^2 D_{22} + D_{12}^2 D_{33} - 2D_{12}D_{13}D_{23}}{D_{23}^2 - D_{22}D_{33}}),$$

$$\chi = u_1\left(D_{33} - \frac{D_{23}^2}{D_{22}}\right),$$

$$\kappa = u_1\left[D_{22} - \frac{(nD_{13} + mD_{22})D_{23}}{nD_{33} + mD_{13}}\right],$$

$$D_{ij} \equiv \partial_i \partial_j (g/T). \tag{31}$$

If we want to calculate thermodynamic quantities at fixed electron density, we can use Newton's method again. Using the compressibility we reach target density after few times of iterations.

## III. $t$-EXPANSION AND $\beta(=1/T)$ EXPANSION

The $t$ expansion of thermodynamic potential up to $t^4$ has been done for single band Hubbard model by Kubo [11] and Liu [12]. For one-dimensional model their expansion becomes

$$-g\beta = \log[\xi] + (\beta t)^2 \frac{2}{\xi^2} G_1 + (\beta t)^4 (\frac{2}{\xi^2} G_2 + \frac{2}{\xi^3} G_3 - \frac{6}{\xi^4} G_1^2) + O((\beta t)^6), \tag{32}$$

where

$$\xi \equiv 1 + 2\cosh(\beta h)e^{\beta A} + e^{\beta(2A-U)},$$

$$G_1 = \cosh(\beta h)e^{\beta A}(1 + e^{\beta(2A-U)}) + \frac{2e^{2\beta A}}{\beta U}(1 - e^{-\beta U}),$$

$$G_2 = \frac{1}{12}\cosh(\beta h)e^{\beta A}(1 + e^{\beta(2A-U)}) + \frac{4e^{2\beta A}}{(\beta U)^2}(1 + e^{-\beta U}) - \frac{8e^{2\beta A}}{(\beta U)^3}(1 - e^{-\beta U}),$$

$$G_3 = \frac{1}{6}\xi e^{\beta A}\cosh(\beta h)(1 + e^{\beta(2A-U)}) + \frac{3e^{2\beta A}}{\beta U}(1 + e^{\beta(2A-U)} - 2\cosh(\beta h)e^{\beta(A-U)})$$

$$+ \frac{2e^{2\beta A}}{(\beta U)^2}(2\cosh(\beta h)e^{\beta A}(1 - 2e^{-\beta U}) - (2 - e^{-\beta U})(1 + e^{\beta(2A-U)})) + \frac{2\xi e^{2\beta A}}{(\beta U)^3}(1 - e^{-\beta U}).$$

$$\tag{33}$$

$G_1$ term is the second order term of $t$ expansion. $G_2$, $G_3$ and $G_1^2$ terms are the fourth order. It is expected that $-g\beta$ is expanded by $\beta t, \beta h, \beta h', \beta u'$, where we put $h' \equiv U/2 - A$, $u' \equiv U/4$:

$$-g\beta = \sum_{n_1,n_2,n_3,n_4 \geq 0} A_{n_1,n_2,n_3,n_4} (\beta t)^{n_1} (\beta h)^{n_2} (\beta h')^{n_3} (\beta u')^{n_4}. \tag{34}$$

From equation (32) and (33) we can calculate all coefficients $A_{n_1,n_2,n_3,n_4}$ at $n_1 < 6$. On the other hand we have

$$-g\beta = \ln 4 + \frac{(\beta t)^2}{2} - \frac{(\beta t)^4}{16} + \frac{(\beta t)^6}{144} - O((\beta t)^8), \qquad (35)$$

when $U = h = h' = 0$. Then we have $A_{6,0,0,0} = 1/144$ and $A_{7,0,0,0} = 0$. In this way, we can determine $A_{n_1,n_2,n_3,n_4}$ at $n_1 + n_2 + n_3 + n_4 \leq 7$ and obtain $\beta$ expansion of the grand potential:

$$\begin{aligned}
g = & -\frac{\ln 4}{\beta} + (h' - u') - \frac{\beta}{4}(2t^2 + 2u'^2 + h^2 + h'^2) - \frac{\beta^2}{4}(h^2 - h'^2)u' \\
& + \frac{\beta^3}{96}\left(6t^4 + 12t^2(h^2 + h'^2) + 32t^2 u'^2 + 8u'^4 + h^4 + 6h^2 h'^2 + h'^4\right) \\
& + \frac{\beta^4}{24}u'(h^2 - h'^2)(6t^2 + 2u'^2 + h^2 + h'^2) \\
& - \frac{\beta^5}{1440}\Big(10t^6 + t^4[306u'^2 + 90(h^2 + h'^2)] + t^2[288u'^4 + 60u'^2(h^2 + h'^2) + 30(h^4 + 6h^2 h'^2 + h'^4)] \\
& + 32u'^6 - 45u'^2(h^4 - 2h^2 h'^2 + h'^4) + h^6 + h'^6 + 15(h^4 h'^2 + h^2 h'^4)\Big) \\
& - \frac{\beta^6}{2880}u'(h^2 - h'^2)\Big(540t^4 + t^2[720u'^2 + 300(h^2 + h'^2)] + 96u'^4 + 40(h^2 + h'^2)u'^2 \\
& + 17(h^4 + h'^4) + 62h^2 h'^2\Big) + O(\beta^7),
\end{aligned} \qquad (36)$$

This expansion up to $\beta^4$ is equivalent to the one obtained by Charret et al. Moreover we could get two more terms higher than theirs in high-temperature expansion of the grand potential. From this expansion we get specific heat and magnetic susceptibility per site at half-filled and zero field case:

$$c = (t^2 + u'^2)\beta^2 - (\frac{3t^4}{4} + 4t^2 u'^2 + u'^4)\beta^4 + (\frac{5t^6}{24} + \frac{51t^4 u'^2}{8} + 6t^2 u'^4 + \frac{2u'^6}{3})\beta^6$$
$$+ O(\beta^8), \qquad (37)$$
$$\chi = \frac{\beta}{2} + \frac{u'\beta^2}{2} - \frac{t^2 \beta^3}{4} - \frac{1}{6}(3t^2 + u'^2)u'\beta^4 + \frac{1}{24}(3t^2 + 2u'^2)t^2\beta^5 + (\frac{3t^4 u'}{8} + \frac{t^2 u'^3}{2} + \frac{u'^5}{15})\beta^6$$
$$+ O(\beta^7). \qquad (38)$$

In figure 1 we plot our numerical results of TBA equations of specific heat at $U = 4, h = 0, A = U/2$. They agree very well with high temperature expansion when $\beta < 0.1$. Dashed line is the numerical calculation of TBA equations by Charret et al. Figure 2 is specific heat at $U = 8$. Figure 3 is the susceptibility at $U = 4, h = 0, A = U/2$.

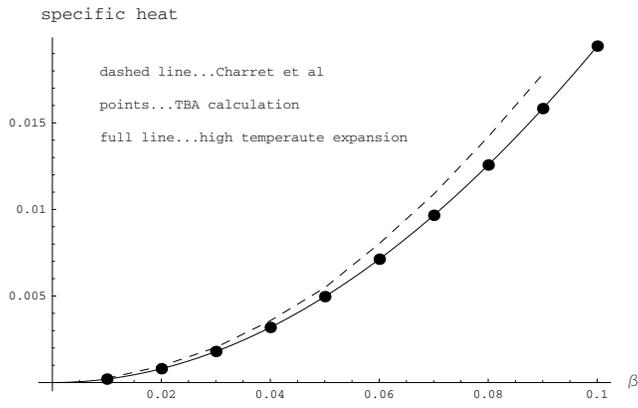

Figure 1

FIG. 1: Specific heat at $U = 4, h = 0, U/2 - A = 0$. Points are our results of TBA calculations and dashed line is Charret et al's calculation. Full line is high temperature expansion.

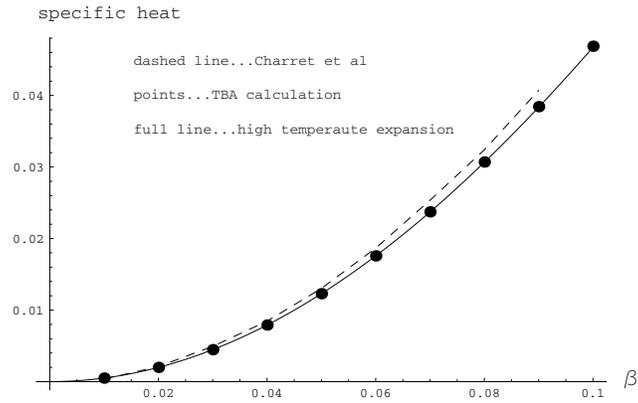

Figure 2

FIG. 2: Specific heat at $U = 8, h = 0, U/2 - A = 0$.

## IV. DISCUSSION AND CONCLUSION

In this paper, we show that TBA equations and QTM formulations by Jüttner et al give completely the same numerical results and coincide also with the high temper-ature expansion. Charret et al's numerical calculation for TBA equations is not correct. Probably the convergence condition is too generous. Our numerical calculation is done by use of mathematica 4.1. Source file is available from http://www.issp.u-tokyo.ac.jp/labs/theory/mtaka/index.html.

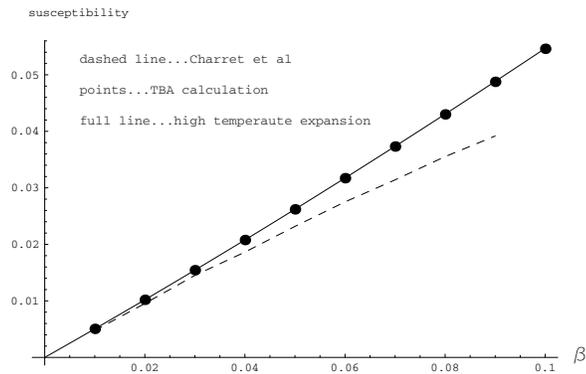

FIG. 3: Magnetic susceptibility at $U = 4, h = 0, U/2 - A = 0$.

Kawakami's calculation and Jüttner's calculation are based on Kepler's method. TBA equations show very slow convergence in Kepler's method. One needs several tens or several hundreds of iterations if one take $n_c = 6$. Jüttner's QTM equation is a bit faster but one needs several tens of iterations. Charret's method seems to be based on Kepler's method as they reported that the TBA equations converged after 2-3 times iterations. This is too short and not reliable. We conclude that the discrepancy between TBA equations with QTM equations or high-temperature expansion comes from their inappropriate convergence conditions.

We get very fast convergence if we adopt Newton's method. After 5-6 times of iterations we get sufficient convergence. Moreover in Newton's method the error decreases with acceleration.

We made numerical programs for TBA and QTM equations. In table I, we observe the dependence of the numerical results on $n_c$. As $n_c$ increases they converge to the high temperature expansion and Jüttner's QTM calculation. Both equations give completely the same numerical results. Especially for the grand potential the values coincide with 5 figures accuracy. We can conclude that both equations are equivalent, although the mathematical equivalence is not yet proved.

About the SO(4) symmetry reference [2] did not treat explicitly. But in reference [1] the symmetry is treated carefully and the same equations are derived. The thermodynamics may not be sensitive to the SO(4) symmetry. In conclusion we can use TBA equations for thermodynamic quantities of 1D Hubbard model. But one needs some numerical technique shown in this paper.

For the XXZ model it is known that the TBA equations and QTM method give the completely the same results numerically [15–18, 20]. Recently, it is shown that TBA equations can be derived from quantum transfer matrix formulations for this model [13, 19]. An intriguing new simple equation, whic has only one unknown function, was also derived both from TBA and QTM [14]. In future, we hope to show the equivalence of two formulations for the Hubbard model.

### Acknowledgments

We aknowlege K. Kubo, N. Kawakami, A. Klümper, V. Korepin, T. Deguchi and A. Ogata for stimulating discussions. This research is supported in part by Grants-in-Aid for the Scientific Research (B) No. 11440103 from the Ministry of Education, Science and Culture, Japan.

|  | $-g$ | $e$ | entropy | $C$ | $\chi$ |
|---|---|---|---|---|---|
| $n_c = 6$ | 14.96245182 | 0.801931104 | 1.37643829 | 0.01943355 | 0.05459858 |
| $n_c = 12$ | 14.96247427 | 0.801886862 | 1.37643611 | 0.01943778 | 0.05465946 |
| $n_c = 24$ | 14.96247646 | 0.801882551 | 1.37643590 | 0.01943818 | 0.05467815 |
| $n_c = 48$ | 14.96247660 | 0.801882261 | 1.37643588 | 0.01943821 | 0.05468324 |
| $n_c = 96$ | 14.96247662 | 0.801882219 | 1.37643588 | 0.01943821 | 0.05468587 |
| $\beta$ exp. | 14.96246886 | 0.801890167 | 1.37643590 | 0.01943825 | 0.05468635 |
| JKS | 14.96246887 | 0.801890163 | 1.37643590 | 0.01943818 | 0.05468632 |

TABLE I: Numerical results of TBA equations for grand potential, energy, entropy, specific heat and susceptibility at $\beta = .1, U = 4, h = U/2 - A = 0$ for various values of $n_c$. We put $L = 64$ and compare with high temperature expansion (36) and Jüttner et al's QTM equations. We find that $n_c = 6$ is practically sufficient.

## APPENDIX A: $t$-EXPANSION FOR TBA EQUATIONS

First, we expand $e_0$ as power series of $U' = 4t/U$,

$$e_0 = -4\left[(\frac{1}{2})^2 \ln 2 U'^{-1} - (\frac{1 \cdot 3}{2 \cdot 4})^2 \frac{\zeta(3)}{3}(1 - \frac{1}{2^2})U'^{-3} + ...\right]. \tag{A1}$$

Then we have $t$ expansion of constant term of $-g\beta$,

$$(A - e_0)\beta = \beta A + \frac{4\ln 2}{\beta U}(\beta T)^2 - \frac{9\zeta(3)}{(\beta U)^3}(\beta t)^4 + O(\beta t)^6. \tag{A2}$$

We put $\Lambda = U'x$. $\sigma_0, \rho_0, \kappa_0$ and $s_1$ are written as follows:

$$\sigma_0(\Lambda) = \frac{t}{U}\text{sech}\frac{\pi}{2}x\left[1 + \frac{\pi^2 t^2}{U^2}(-1 + 2\tanh^2\frac{\pi}{2}x) + O(t^4)\right],$$
$$\rho_0(k) = \frac{1}{2\pi} + \frac{2t\ln 2}{\pi U}\cos k + O(t^3),$$
$$\kappa_0(k) = -2t\cos k - \frac{U}{2} - \frac{4t^2 \ln 2}{U} + O(t^3),$$
$$s_1(\Lambda) = -\frac{2\pi t^2}{U}\text{sech}\frac{\pi x}{2} + O(t^3). \tag{A3}$$

Integral $\int_{-\infty}^{\infty} d\Lambda s(\Lambda - \sin k)...$ becomes

$$\int_{-\infty}^{\infty} dx \bar{s}(x)\left[1 + \frac{2\pi t}{U}\sin k \tanh\frac{\pi}{2}x + (\frac{2\pi t}{U}\sin k)^2(-\frac{1}{2} + \tanh^2\frac{\pi}{2}x) + O(t^3)\right],$$

where
$$\overline{s}(x) \equiv \frac{1}{4}\text{sech}\frac{\pi}{2}x.$$

TBA equations are

$$\ln \zeta(k) = \beta(-\frac{U}{2} - 2t\cos k - \frac{4t^2 \ln 2}{U}) + \int_{-\infty}^{\infty} dx \overline{s}(x)$$
$$\times \left[1 + \frac{2\pi t}{U}\sin k \tanh\frac{\pi}{2}x + (\frac{2\pi t}{U}\sin k)^2(-\frac{1}{2} + \tanh^2\frac{\pi}{2}x)\right] \ln \frac{1 + \eta_1'(x)}{1 + \eta_1(x)},$$
$$\ln \eta_1(x) = -\frac{8\pi\beta t^2}{U}\overline{s}(x) - \frac{4t}{U}\int_{-\pi}^{\pi} \overline{s}(x)(1 + \frac{2\pi t}{U}\sin k \tanh\frac{\pi}{2}x)\ln(1 + \zeta(k))\cos k dk$$
$$+ \overline{s} * \ln(1 + \eta_2(x)),$$
$$\ln \eta_1'(x) = -\frac{4t}{U}\int_{-\pi}^{\pi} \overline{s}(x)(1 + \frac{2\pi t}{U}\sin k \tanh\frac{\pi}{2}x)\ln(1 + \zeta(k))\cos k dk + \overline{s} * \ln(1 + \eta_2'(x)),$$
$$\ln \eta_j(x) = \overline{s} * \ln[(1 + \eta_{j-1}(x))(1 + \eta_{j+1}(x))],$$
$$\ln \eta_j'(x) = \overline{s} * \ln[(1 + \eta_{j-1}'(x))(1 + \eta_{j+1}'(x))]. \tag{A4}$$

$-g\beta$ is given by

$$-g\beta = \beta A + \frac{4\ln 2}{\beta U}(\beta t)^2 + \int_{-\pi}^{\pi} (\frac{1}{2\pi} + \frac{2\ln 2t}{\pi U}\cos k)\ln(1 + \zeta(k))dk$$
$$+ \int_{-\infty}^{\infty} \overline{s}(x)(1 + \frac{\pi^2 t^2}{U^2}(-1 + 2\tanh\frac{\pi}{2}x))\ln(1 + \eta_1(x))dx. \tag{A5}$$

We expand fuctions by power series of $\beta t$,

$$\ln(1 + \eta_j(x)) = \ln \frac{\alpha_j}{\alpha_j - 1} + (\beta t) f_j^{(1)}(x) + (\beta t)^2 f_j^{(2)}(x) + ...,$$
$$\ln(1 + \eta_j'(x)) = \ln \frac{\alpha_j'}{\alpha_j' - 1} + (\beta t) f_j'^{(1)}(x) + (\beta t)^2 f_j'^{(2)}(x) + ...,$$
$$\ln(1 + \zeta(k)) = \ln \frac{z_0}{z_0 - 1} + (\beta t) z^{(1)}(k) + (\beta t)^2 z^{(2)}(k) + .... \tag{A6}$$

We get

$$\ln \eta_j(x) = \ln \frac{1}{\alpha_j - 1} + (\beta t)\alpha_j f_j^{(1)}(x) + (\beta t)^2[\alpha_j f_j^{(2)}(x) + \frac{1}{2}(\alpha_j - \alpha_j^2)(f_j^{(1)}(x))^2] + ...,$$
$$\ln \eta_j'(x) = \ln \frac{1}{\alpha_j' - 1} + (\beta t)\alpha_j' f_j'^{(1)}(x) + (\beta t)^2[\alpha_j' f_j'^{(2)}(x) + \frac{1}{2}(\alpha_j' - \alpha_j'^2)(f_j'^{(1)}(x))^2] + ...,$$
$$\ln \zeta(k) = \ln \frac{1}{z_0 - 1} + (\beta t) z_0 z^{(1)}(k) + (\beta t)^2[z_0 z^{(2)}(k) + \frac{1}{2}(z_0 - z_0^2)(z^{(1)}(k))^2] + ..., \tag{A7}$$

where

$$\alpha_j = \frac{[j+1]^2}{[j][j+2]}, \quad \alpha_j' = \frac{[j+1]'^2}{[j]'[j+2]'}, \quad z_0 = 1 + e^{\beta U/2}\frac{[2]}{[2]'}. \tag{8}$$

and $[j]$ and $[j]'$ are $q$-integers defined by

$$[j] \equiv \frac{\sinh j\beta h}{\sinh \beta h}, \quad [j]' \equiv \frac{\sinh j\beta(U/2 - A)}{\sinh \beta(U/2 - A)}. \tag{9}$$

We have

$$f_j^{(1)}(x) = f_j'^{(1)}(x) = 0, \quad z^{(1)}(k) = -2z_0^{-1} \cos k. \tag{10}$$

Equations for $f_j^{(1)}(x), f_j'^{(1)}(x), z^{(2)}(k)$ are

$$\alpha_1 f_1^{(2)}(x) = -\frac{8\pi}{\beta U}(1 - z_0^{-1})\overline{s}(x) + \overline{s} * f_2^{(2)}(x),$$

$$\alpha_1' f_1'^{(2)}(x) = \frac{8\pi}{\beta U} z_0^{-1} \overline{s}(x) + \overline{s} * f_2'^{(2)}(x),$$

$$\alpha_j f_j^{(2)}(x) = \overline{s} * (f_{j-1}^{(2)}(x) + f_{j+1}^{(2)}(x)),$$

$$\alpha_j' f_j'^{(2)}(x) = \overline{s} * (f_{j-1}'^{(2)}(x) + f_{j+1}'^{(2)}(x)),$$

$$z_0 z^{(2)}(k) = 2(1 - z_0^{-1})\cos^2 k - \frac{4\ln 2}{\beta U}$$

$$+ \int_{-\infty}^{\infty} dx \overline{s}(x)(f_1'^{(2)}(x) - f_1^{(2)}(x)). \tag{11}$$

Same kind of equations have appeared in high temperature expansion of XXX model. See p.124-126 of [1] or [15]. Solutions of these equations are

$$f_j^{(2)}(x) = -\frac{8\pi}{\beta U} \frac{(1 - z_0^{-1})}{[2][j+1]}\Big([j+2]\overline{a}_j(x) - [j]\overline{a}_{j+2}(x)\Big),$$

$$f_j'^{(2)}(x) = \frac{8\pi}{\beta U} \frac{z_0^{-1}}{[2]'[j+1]'}\Big([j+2]'\overline{a}_j(x) - [j]'\overline{a}_{j+2}(x)\Big),$$

$$z^{(2)}(k) = \frac{2(z_0 - 1)}{z_0^2}\Big(\cos^2 k - \frac{1}{\beta U[2]^2}(1 + \frac{[2]}{[2]'}e^{-\beta U/2})\Big),$$

$$\overline{a}_j(x) \equiv \frac{j}{\pi(x^2 + j^2)}. \tag{12}$$

Substituting (10) and (12) into (A6) and (A5) we get

$$-g\beta = \beta A - \ln(1 - z_0^{-1}) + \ln(2\cosh \beta h)$$

$$+ (\beta t)^2 \Big\{ \frac{4\ln 2}{\beta U} + \frac{1}{2\pi} \int_{-\pi}^{\pi} [z^{(2)}(k) + \frac{4\ln 2}{\beta U} \cos k z^{(1)}(k)] dk$$

$$+ \int_{-\infty}^{\infty} \overline{s}(x)[f^{(2)}(x) + \ln\frac{\alpha_1}{\alpha_1 - 1}\frac{\pi^2 t^2}{\beta^2 U^2}(-1 + 2\tanh^2 \frac{\pi}{2}x)]dx \Big\}$$

$$= \ln \xi + \frac{2(\beta t)^2}{\xi^2}[\cosh(\beta h)e^{\beta A}(1 + e^{\beta(2A-U)}) + \frac{2}{\beta U}e^{2\beta A}(1 - e^{-\beta U})], \tag{13}$$

where $\xi$ is defined in (33). This coincides with (32) up to the second order. Thus we have proven that TBA equations give correct $t$ expansion of $-g\beta$ up to $(t\beta)^2$.